\documentclass[aps,prl,prabib,twocolumn,showpacs,nofootinbib]{revtex4}
\usepackage{graphicx} \usepackage{amsmath} \usepackage{amssymb}
\usepackage{amsfonts} \usepackage{bm}

\begin{document}

\newcommand{\be}{\begin{equation}} \newcommand{\ee}{\end{equation}}
\newcommand{\bea}{\begin{eqnarray}}\newcommand{\eea}{\end{eqnarray}}

\title{Scaling anomaly in cosmic string background}

\author{Pulak Ranjan Giri} \email{pulakranjan.giri@saha.ac.in}

\affiliation{ Saha Institute of Nuclear Physics, 1/AF Bidhannagar, Calcutta
700064, India}

\begin{abstract}
We show that the classical scale symmetry of a particle moving in
cosmic string background is broken upon inequivalent quantization of
the classical system, leading to anomaly. The consequence of this
anomaly is the formation of single bound state in the coupling
interval $\gamma\in(-1,1)$.  The inequivalent quantization is
characterized by a 1-parameter family of self-adjoint extension
parameter $\omega$. It has been conjectured that the formation of
loosely bound state in cosmic string background may lead to the so
called  anomalous scattering cross section for the particles, which
is usually seen in molecular physics.
\end{abstract}



\date{\today}

\maketitle

Symmetry and corresponding symmetry breaking \cite{peskin} is an extremely
important issue in physics because of their consequences in different physical
processes. Usually in physics we consider three different sorts of symmetry
breaking, i.e., spontaneous, explicit and anomalous symmetry
breaking. However, anomalous symmetry breaking \cite{trei,dono} occurs when
some kind of classical invariance of a system is violated upon quantization.
Physical realization of anomaly in nature has been observed  in high energy
physics since the introduction of chiral anomaly \cite{bell,adler}. In the
study of elementary particles in standard model the concept of anomaly has
been used successfully \cite{dono,epele} and in beyond standard model and
string theory \cite{pol} as well. Besides  its phenomenological importance,
the study of anomaly has become the subject of many theoretical
investigations starting from molecular physics  \cite{camblong} to black hole 
\cite{govind}. 
For example, in path integral formulation, chiral anomaly is
due to lack of invariance of the functional measure.  In quantum mechanics  an
operator becomes anomalous when it does not keep the domain of the Hamiltonian
invariant. There is another interesting example of anomaly, which occurs in
molecular physics \cite{camblong} in quantum mechanical context. For example,
interaction of an electron in the field of a polar molecule is a simple
example of anomaly, where the classical scaling symmetry of the system is
broken once it goes inequivalent quantization \cite{giri}. The obvious
consequence of this  scaling anomaly in molecular physics is the occurrence of
bound state and the the dependence of momentum in the phase shift of 
scattering cross section.

The problem of quantum anomaly which we consider here may occur in cosmic
string background, when a particle is moving in it. This problem has received
lots of interest due to its analogy  \cite{pes}  with Aharanov-Bohm effect
\cite{aha}. In  relativistic theory it has been shown \cite{sousa} that the 
Dirac
equation in cosmic string background needs nontrivial boundary condition to be
imposed on the spinor wave-function at the origin. In language of mathematics
the construction of nontrivial boundary condition  is usually  called
self-adjoint extensions \cite{reed}. The extensions can be characterized by
independent parameters and different value of parameters lead to inequivalent
theories. It has been observed \cite{wilczek} that in cosmic string scenario
the fermionic charge can be non-integral multiple of Higgs charge. Since the
flux is quantized with respect to the Higgs charge, it will lead to nontrivial
Aharanov-Bohm scattering of fermion. This result has Phenomenological
importance because, the cross section is much larger than the one coming from
gravitational scattering.
In this letter we will discuss about another possible enhancement of
scattering cross section due to the temporaty formation of loosely bound state
in cosmic string background.

In non-relativistic theory\cite{fil}, the consideration of
inequivalent quantization is also inevitable in order to get bound
state for the particle moving in cosmic string background. In Ref.
\cite{ger1}  gravitational scattering by particles of a spinning
source in two dimension has been studied. There, it  has been shown
that the energy eigenvalue and corresponding eigenfunction of a
particle in the field of a massless spinning source is equivalent to
that in a background Aharanov-Bohm gauge field of an infinitely thin
flux tube. This topological effect also appears in elastic solids
\cite{aze}. The anomaly which we will discuss here has the
consequences, which is known for quite a some time. But,
surprisingly the issue of anomaly has remained unnoticed as far as
we know. Most importantly, despite the  similarity with electron
polar molecule system, there is no discussion in literature about
the possible anomaly in scattering cross section, which is usually seen  in
molecular physics \cite{giri}.

This letter has been organized in the following way: First, we study the
scaling symmetry of the classical system, which undergoes anomalous symmetry
breaking
upon quantization; Second, we made an inequivalent quantization of the system,
which is responsible for anomaly and discuss its consequences; Third, we draw
an analogy with molecular physics and conjecture that in cosmic string
background there may have anomalous scattering cross section due to the
loosely bound state.

First, what is scaling symmetry and why it is important in cosmic string
background? Scale transformation can be defined by the transformation
$\bf{r}\to\beta\bf{r}$ and $t\to\beta^2t$, where $\beta$ is the scaling
factor. In classical physics when the action is invariant under this
transformation, then the corresponding system has scale symmetry. since in
non-relativistic quantum theory, cosmic string induces a $1/r^2$ potential to
the the particle moving in its background, the relevant classical symmetry
would be the scale symmetry. To be more specific, the Hamiltonian for the
system $H=\frac{{\bf p}^2}{2M}$, scales as $H\to \frac{1}{\beta^2}H$. The
scale invariance of this Hamiltonian means, if $\psi$ is an eigenstate of the
Hamiltonian $H$ with eigenvalue $E$, i.e., $H\psi= E\psi$, the
$\psi_\beta=\psi(\beta \bf r)$ will also be an eigenstate of the same
Hamiltonian with energy $E/\beta^2$. This essentially means that the system
with scale symmetry does not have any lower bound in energy; that means it
cannot have any bound state. Scale symmetry is however a part of larger
conformal symmetry formed  by three generators: the Hamiltonian $H$, the
Dilatation generator $D= tH- \frac{1}{4}(\bf{r}.\bf{p}+ \bf{p}.\bf{r})$ and
the conformal generator $K= Ht^2-\frac{1}{2}(\bf{r}.\bf{p}+ \bf{p}.\bf{r}) +
$$\frac{1}{2}M\bf{r}^2$. They form the $SO(2,1)$ algebra:
$[D,H]= -i\hbar H$, $[D,K]= i\hbar K$, $[H,K]= 2i\hbar D$ \cite{wyb}.
 In cosmic string case
surprisingly this scale symmetry has not been noticed so far.

Second, we consider a non-relativistic particle of mass $M$, moving in the 
background
of cosmic string. The background is described by the space-time metric
in cylindrical coordinate $(r,\phi,z)$ as
\begin{eqnarray}
ds^2 = dt^2- dz^2- dr^2 - \alpha^2r^2d\phi^2\,,
\end{eqnarray}
where $\alpha=1- 4G\mu <1 $ characterizes the string, with $\mu$ is the mass
per unit length of the string, $G$ gravitational constant. The constant $\alpha$ introduces an angular
deficit of $2\pi(1-\alpha)$ in the Minkowski space-time and is responsible for
inducing scale invariant $1/r^2$ potential in non-relativistic quantum
system. Due to cylindrical symmetry of the space, we can easily see that the
motion of the particle in the z direction is basically a free particle motion,
described by the wave-function $e^{ikz}$. $k$ is wave-vector of the particle
along the $z$ direction. Since we are considering an infinite cosmic string
along the $z$ direction, it is enough to discuss the motion of the particle on
the plane perpendicular to the $z$ direction. The motion of the particle on
the plane perpendicular to the $z$ axis is described by the time independent
Schr\"{o}dinger equation
\begin{eqnarray}
-\frac{\hbar^2}{2M}\left[\frac{1}{r}\frac{\partial}{\partial
r}(r\frac{\partial}{\partial r}) + \frac{1}{\alpha^2
r^2}\frac{\partial^2}{\partial\phi^2}\right]\Psi= E\Psi
\label{schrodinger}
\end{eqnarray}
Exploiting  the periodicity condition  \cite{kow}
\begin{eqnarray}
\Psi(\phi+2\pi)= e^{2\pi\lambda i}\Psi(\phi)
\label{periodicity}
\end{eqnarray}
where $\lambda\in[0,1)$, the wave-function can be separated as $\Psi(r,\phi) =
R(r)e^{i(m+\lambda)\phi}$ and (\ref{schrodinger}) gives the radial
equation
\begin{eqnarray}
 H_r R(r) \equiv -\left [ \frac{d^2}{dr^2} + \frac{1}{r}\frac{d}{dr}
+\frac{\gamma^2}{r^2} \right ]R(r) = \mathcal E R(r), \label{radial}
\end{eqnarray}
where $H_r$ is the radial Hamiltonian, with eigenvalue $\mathcal E =
\frac{2M E}{\hbar^2}$, $\gamma = \frac{m+\lambda}{\alpha}$ and
$m=0,\pm 1, \pm 2, \cdots$. We will now discuss the solution of the
Hamiltonian $H_r$. To discuss that we need to know some general
property of an operator, let say $\mathcal O$. For the moment let us
restrict ourself to the case of unbounded operator, because the
Hamiltonian we are discussing is unbounded. Now, it is known that
for an unbounded operator $\mathcal O$, we first need to define the
domain $D(\mathcal O)$.  We are  interested in operators that are
defined on  a dense domain in the Hilbert space. This allows us to
construct the adjoint operator  $\mathcal{O}^*$ and the
corresponding domain $D(\mathcal {O}^*)$.  By definition, $\mathcal O$
is self-adjoint if and only if $D(\mathcal O)= D(\mathcal{O}^*)$. A
better way of saying  this is looking at the deficiency indices,
which are defined as follows. Let $\mathcal K_\pm= Ker(i\pm
\mathcal{O}^*)$, where $Ker(X)$ is the kernel of the operator. The
deficiency indices $n_\pm$ are the dimension of the kernel
$\mathcal {K}_\pm$. If $n_\pm=0$, then the operator  $\mathcal O$ is
essentially self-adjoint. If $n_+=n_-=n\neq 0$, then $\mathcal O$ is
not self-adjoint but admits  self-adjoint extensions. Different
self-adjoint extensions of the operator are in one-one
correspondence with unitary maps from $\mathcal K_+$ to $\mathcal
K_-$, that is  labeled by a $U(n)$ matrix. Finally, if $n_+\neq
n_-$, then the operator  $\mathcal O$ cannot be made self-adjoint.
\begin{figure}
\includegraphics[width=0.4\textwidth, height=0.2\textheight]{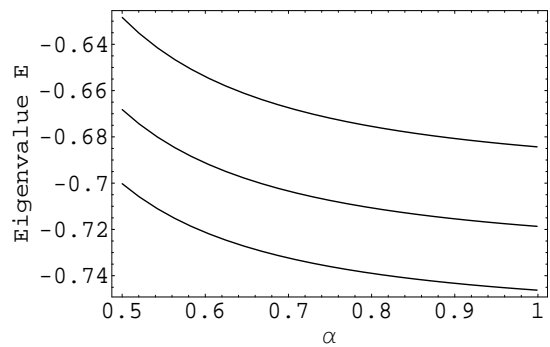}
\caption {A plot of bound state  energy $E$ (in $\frac{\hbar^2}{2M}$
unit) of particle as a function of  $\alpha= 1-4\mu G$ of the cosmic
string for three different values of the self-adjoint extension
parameter $\omega$ and with $m=0, \lambda=1/2$. From top to bottom
$\omega= \frac{\pi}{7}, \frac{\pi}{8},\frac{\pi}{9}$ respectively.}
\end{figure}

Let us now come back to the discussion of our operator of interest, which is
$H_r$.
The Hamiltonian $H_r$ is defined on the domain $\mathcal{L}^2[R^+,
rdr]$. Classically, this system is scale invariant, because the coupling
constant of the inverse square potential $\gamma$ is dimensionless
constant. However, quantum mechanical analysis of this operator is much more
subtle. The Hamiltonian $H_r$ is essentially self-adjoint only for
$\gamma^2\geq 1$, the domain of the Hamiltonian is
\begin{eqnarray}
\mathcal{D}_0=\{\psi\in \mathcal{L}^2(rdr), \psi(0)=\psi'(0)=0\}
\end{eqnarray}
For $\gamma\in(-1,1)$, the Hamiltonian is not essentially
self-adjoint and therefore cannot play a role for the Hamiltonian
and so has to be extended to another operator. Note that only $m=0,
\lambda\in[0,1)$ and $m=-1, \lambda\in(0,1)$ belong to the interval
$\gamma\in(-1,1)$. For this case the deficiency indices are $(1,1)$,
and so the self-adjoint extensions are labeled by a $U(1)$ parameter
$e^{i\omega}$, which labels the domain $\mathcal {D}_\omega$ of the
Hamiltonian $H_\omega$. The set $\mathcal {D}_\omega$ contains all
the vectors of the form $\phi_+ + e^{i\omega}\phi_-$ together with
the element of the domain $\mathcal {D}_0$. The 
solutions $\phi_\pm$ are
\begin{eqnarray}
\phi_\pm = K_\gamma(re^{\mp i\pi/4})\,,
\end{eqnarray}
where $K_\gamma$ is the modified Bessel function \cite{abr}. The
behavior near singularity  $r\to 0$  of $\phi_+ + e^{i\omega}\phi_-$
is
\begin{eqnarray}
\phi_+ + e^{i\omega}\phi_-\simeq \mathcal{A}_+\left(\frac{r}{2}\right)^\gamma
+ \mathcal{A}_-\left(\frac{2}{r}\right)^{\gamma}
\end{eqnarray}
where, $\mathcal{A}_\pm= -\frac{\pi i}{\sin(\pi\gamma)}
\frac{\cos(\frac{\omega}{2}\pm \frac{\pi\gamma}{4})}{\Gamma(1 \pm\gamma)} $

We can now solve the eigenvalue problem (\ref{radial}). For $\gamma^2\geq 1$
there are no bound states. More precisely there are no normalizable  solution
of the Schr\"{o}dinger equation with negative energy. However, for
$\gamma\in(-1,1)$, there is exactly one bound state with energy $\mathcal E$,
where
\begin{eqnarray}
\mathcal E= -\left[\frac{\cos\frac{1}{4}\left(2\omega+ \gamma\pi\right)}
{\cos\frac{1}{4}\left(2\omega- \gamma\pi\right)}\right]^{\frac{1}{\gamma}}
\label{bound1}
\end{eqnarray}
and the corresponding eigenfunction is
\begin{eqnarray}
R(r)= K_\gamma(\sqrt{\mathcal{E}} r)
\label{evalue1}
\end{eqnarray}
Here we make some observation regarding our bound state solution
(\ref{bound1}) and (\ref{evalue1}). Note that imposing time-
reversal symmetry in (\ref{periodicity}) we get \cite{kow}
\begin{eqnarray}
\hat T\Psi(\phi+2\pi)= e^{-2\pi\lambda i}\hat T\Psi(\phi)\,.
\label{timerevarsal}
\end{eqnarray}
The consistency of (\ref{periodicity}) and (\ref{timerevarsal})
demands that either $\lambda=0$ or $\lambda=1/2$. However,
$\lambda=0$ is not interesting because, in order to keep $\gamma=
\frac{m+\lambda}{\alpha}$  in the interval $\gamma\in(-1,1)$ for
$\lambda=0$, we have $m=0$. This makes the system independent of the
effect of cosmic string characterized by $\alpha$. So the
interesting case for our purpose are (1) $\lambda=1/2$, $m=0$ and
$\alpha\in(1/2,1)$  and (2) $\lambda=1/2$, $m=-1$ and
$\alpha\in(1/2,1)$.
The energy eigenvalue for $m=0, \lambda=1/2$  has been plotted in
FIG. 1. For $m=0, \lambda= - 1/2$, the plot will be same, because
the eigenvalue (\ref{evalue1}) is symmetric with respect to
$\gamma$.  From FIG. 1, it can be seen that there may exist bound
state for arbitrarily small value of the mass density $\mu$ for the
cosmic string.

The existence of bound state is in contradiction with the scale
invariance, since scale invariance implies that there is no length
scale in the problem, whereas the existence of bound state provides a
scale. This  can be resolved by looking at how scaling is
implemented in the quantum theory.  The scaling operator is
\begin{eqnarray}
\Lambda= \frac{rp + pr}{2}
\end{eqnarray}
where $p= -i\frac{d}{dr}$. It is easily seen that $\Lambda$ is symmetric on
the domain $\mathcal {D}_0$ of $H_r$, and that for $\gamma^2\geq 1$, $\Lambda$
leaves invariant the domain of the Hamiltonian. For $\gamma^2\in(-1,1)$,
\begin{eqnarray}
\Lambda\phi= -i\left(\phi + 2r\phi'\right)\,,
\end{eqnarray}
where $\phi$ is any element, belonging to the domain $\mathcal{D}_\omega$. The
small $r$ behavior of the function $\Lambda\phi$ is of the form
\begin{eqnarray}
\nonumber\Lambda\phi \simeq -\frac{i}{2}
\left[(1+2\gamma)\mathcal{A}_+\left(\frac{r}{2}\right)^\gamma  +
(1-2\gamma)\mathcal{A}_-\left(\frac{2}{r}\right)^{\gamma}\right]
\end{eqnarray}
where, the constants $\mathcal{A}_\pm$ are defined above.

So, $\Lambda\phi$ clearly does not leave the domain of the
Hamiltonian invariant. Scale invariance is thus anomalously broken,
and this breaking occur precisely when the Hamiltonian admits
nontrivial self-adjoint extensions. This also explains the quantum
mechanical emergence of a length scale, namely the bound state
energy. We must remark here that there does exist self-adjoint
extensions that preserve scale invariance. For example, for $\omega=
(1\pm\frac{\gamma}{2})\pi$ there does not exist any bound state. From
the point of view of the domains, the operator $\Lambda$ leaves the
domain invariant, implying that the scaling can be consistently
implemented in the quantum theory.

Third, we  come to the question of anomalous scattering cross section.
We discuss this anomalous scattering in analogy with molecular
physics \cite{giri,ger,hurst1,turner2}.  In molecular physics,
electron moving in the dipole field of a molecule experiences
inverse square potential and electrons are loosely captured by this
inverse square potential. It has been observed that the experimental
scattering cross section for the electron is much larger than the
theoretically calculated value. It is usually  argued \cite{note1}
that the observed discrepancy between the experimentally observed
and theoretically calculated scattering cross section of electrons
is due to the formation loosely bound state in the inverse square
potential. In our case also the situation is exactly same. The
particle experiences the same attractive inverse square potential
apart from the usual centrifugal term, while moving in the
background metric of cosmic string. In this letter, it has been
shown that particle can form loosely bound state in cosmic string
background, see FIG. 1  for the behavior of bound state energy with
respect to the constant $\alpha$, characteristic of the string. So
in same line with molecular physics we can conjecture that the
formation of loosely bound state in cosmic string background may
lead to the anomalous scattering cross section for the particles.
However, the crucial difference between the molecular physics and
our analysis is that in molecular physics the scaling symmetry
explicitly breaks down in reality due to the finite size of the
dipole but in case of cosmic string there seems to be no such
explicit scaling symmetry breaking in non-relativistic theory.

In conclusion, we have shown the existence of scaling
anomaly in cosmic string background. The consequence of this anomaly
is the existence of bound state of the particle moving in cosmic
string background. We have conjectured in analogy with molecular
physics that there may have anomalous scattering due to the loosely
bound state of the particle in cosmic string background. Although
the conjecture is based on purely qualitative basis at this stage,
it is however an interesting issue to  study.


\begin{thebibliography}{99}
\bibitem{note1} Anomalous scattering cross section for electron in the
field of polar molecule due to temporary capture of the electron by
the polar molecule was first pointed out by Turner. See the paper of
Turner in reference \cite{ger} for detail discussion.
\end{thebibliography}
\end{document}